\documentclass[conference]{IEEEtran}
\IEEEoverridecommandlockouts
\usepackage{cite}
\usepackage{amsmath,amssymb,amsfonts}
\usepackage{graphicx}
\usepackage{textcomp}
\usepackage{xcolor}
\usepackage{color,array}
\usepackage{orcidlink}
\usepackage{tcolorbox}
\usepackage{booktabs}
\usepackage{float}
\usepackage{amsmath} 

\def\BibTeX{{\rm B\kern-.05em{\sc i\kern-.025em b}\kern-.08em
    T\kern-.1667em\lower.7ex\hbox{E}\kern-.125emX}}

\usepackage[inkscapelatex=false]{svg}
\svgpath{{./}}
\usepackage{amsmath}
\usepackage{booktabs}
\usepackage{amssymb}
\usepackage{amsmath}
\usepackage{multirow}
\usepackage{float} 
\usepackage{cite}
\usepackage{enumitem}
\usepackage[inkscapelatex=false]{svg}
\svgpath{{./}} % Path to your SVG files (adjust if your SVGs are in a subdirectory)
\usepackage{tcolorbox}
\usepackage{booktabs}
\usepackage{multirow}
\usepackage{array}

\usepackage{amsmath,amssymb,amsfonts}
\usepackage{algpseudocode} 
\usepackage{algorithm}
\usepackage{graphicx}
\usepackage{subcaption}
\usepackage{textcomp}
\usepackage{xcolor}
\usepackage{titlesec}
\usepackage{orcidlink}
\usepackage{amsmath}
\usepackage{algorithm}
\usepackage{algorithmicx}
\usepackage{algpseudocode}
\begin{document}

%Predicting Lead Risk Without Testing: A Data-Driven Approach Using Health Coverage Patterns
% Can LLMs Analyze Structured Health Data? A Case Study with Lead Prevalence in Children

\title{ Can LLMs Help Allocate Public Health Resources? A Case Study on Childhood Lead Testing\\}

%\thanks{\textsuperscript{$\mathsection$} J. Chen acknowledged the support by the Fordham Office of Research through a Fordham AI Research (FAIR) Grant. }

\author{Mohamed Afane\textsuperscript{1,\orcidlink{0009-0005-8959-6895}}, Ying Wang\textsuperscript{2,\orcidlink{0000-0002-9004-7253}}, and Juntao Chen\textsuperscript{1*,\orcidlink{0000-0001-7726-4926}}\\
\small \textsuperscript{1}Department of Computer and Information Sciences, Fordham University, New York, United States\\
\small \textsuperscript{2}Department of Systems Engineering, Stevens Institute of Technology, Hoboken, New Jersey, United States\\
\small Email: \{mafane, jchen504\}@fordham.edu, ywang6@stevens.edu
\thanks{J. Chen acknowledges the support through a grant from The Fordham Center for Community Engaged Learning (CCEL).
}}

\maketitle

\begin{abstract}
Public health agencies face critical challenges in identifying high-risk neighborhoods for childhood lead exposure with limited resources for outreach and intervention programs. To address this, we develop a Priority Score integrating untested children proportions, elevated blood lead prevalence, and public health coverage patterns to support optimized resource allocation decisions across 136 neighborhoods in Chicago, New York City, and Washington, D.C. We leverage these allocation tasks, which require integrating multiple vulnerability indicators and interpreting empirical evidence, to evaluate whether large language models (LLMs) with agentic reasoning and deep research capabilities can effectively allocate public health resources when presented with structured allocation scenarios. LLMs were tasked with distributing 1,000 test kits within each city based on neighborhood vulnerability indicators. Results reveal significant limitations: LLMs frequently overlooked neighborhoods with highest lead prevalence and largest proportions of untested children, such as West Englewood in Chicago, while allocating disproportionate resources to lower-priority areas like Hunts Point in New York City. Overall accuracy averaged 0.46, reaching a maximum of 0.66 with ChatGPT 5 Deep Research. Despite their marketed deep research capabilities, LLMs struggled with fundamental limitations in information retrieval and evidence-based reasoning, frequently citing outdated data and allowing non-empirical narratives about neighborhood conditions to override quantitative vulnerability indicators.
\end{abstract}

\begin{IEEEkeywords}
Childhood Lead Exposure, Public Health Resource Allocation, Large Language Models
\end{IEEEkeywords}

\section{Introduction}

Lead poisoning accounts for an estimated \$50.9 billion in annual economic losses in the United States due to cognitive impairments \cite{hauptmanUpdateChildhoodLead2017}, but the profound human costs extend far beyond financial metrics. Children are particularly vulnerable due to their developing nervous systems and higher rates of lead absorption \cite{hauptmanUpdateChildhoodLead2017, luo2012effects, finkelstein1998low}, facing severe consequences including neurodevelopmental damage and reduced academic performance \cite{canfieldIntellectualImpairmentChildren2003}, with effects that persist throughout their lives and into adulthood. These challenges are especially acute in urban settings, where aging buildings, deteriorating water infrastructure, and other environmental hazards contribute to heightened exposure risks \cite{jacobsPrevalenceLeadbasedPaint2002, bordenVulnerabilityUSCities2007, moodyLeadEmissionsPopulation2017, mielke1999lead}.

Numerous studies have demonstrated that lead exposure and testing inequities disproportionately affect urban neighborhoods in U.S. cities, with some communities facing significantly higher vulnerability and fewer resources to address the issue \cite{afaneAnalyzingOptimizingDistribution2025,osheaLeadPollutionDemographics2021, moodyRacialGapChildhood2016}. These disparities reflect broader systemic challenges in ensuring equitable access to testing and remediation, particularly in historically marginalized and economically disadvantaged areas where environmental hazards and social inequities often converge. On a global scale, similar patterns are observed, with marginalized communities often bearing the greatest burden of lead exposure due to limited resources and infrastructure \cite{wangBloodLeadLevels2006,olympioWhatAreBlood2017,kimHowDoesLow2018}. Multiple factors have been identified as significant contributors to elevated blood lead levels (BLL), including socioeconomic status \cite{gleasonDrinkingWaterLead2019}, racial demographics \cite{moodyRacialGapChildhood2016}, environmental hazards such as water quality \cite{pocockEffectsTapWater1983}, and housing conditions \cite{spanierContributionHousingRenovation2013}. However, this study focuses on a critical yet under-explored factor that demonstrates even stronger correlations with lead vulnerability: the type of health coverage available in these communities.

\begin{figure*}[t]
   \centering
     \includegraphics[width=0.95\textwidth]{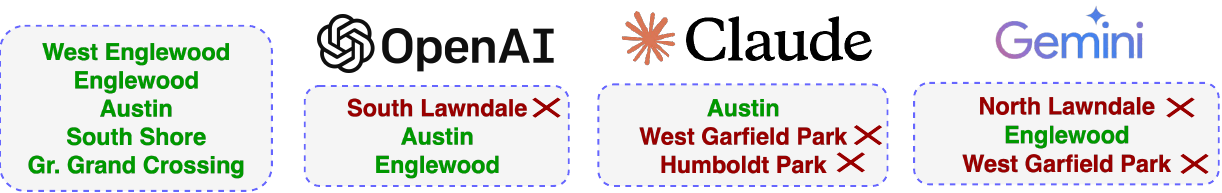}
   \caption{Comparison of LLM resource allocation recommendations for lead testing kits against empirical vulnerability rankings. The left panel shows the five highest-priority neighborhoods based on historic data and vulnerability metrics. The right panels display top-three allocations from ChatGPT, Claude, and Gemini for Chicago. Green highlights indicate correct identification of high-priority areas, while red highlights with crosses mark lower-priority neighborhoods incorrectly placed in the top three.}
   \label{fig:v1}
\end{figure*}

In the United States, health insurance coverage falls into two primary categories: private and public. Private coverage, which includes employer-based insurance and direct-purchase plans, covers approximately 66\% of the population \cite{bureauHealthInsuranceCoverage}. Public coverage primarily consists of Medicaid, which serves low-income individuals and families, and Medicare, which covers adults aged 65 and older \cite{altmanMedicareMedicaid502015}. As of 2024, public coverage reaches approximately 35.5\% of Americans, with Medicaid accounting for 17.6\% of total coverage \cite{bureauHealthInsuranceCoverage}. For children under 19, public coverage rates are notably higher at 34.2\%, reflecting the program's focus on vulnerable populations. These coverage types differ substantially in their access to preventive care, diagnostic services, and health outcomes, making insurance type a potentially valuable indicator of broader health disparities. Previous research suggests that health insurance access significantly impacts health outcomes, with uninsured individuals facing heightened risks of chronic conditions. Public health coverage, though essential for reducing medical costs, has been linked to a higher prevalence of health issues such as elevated blood pressure compared to private health coverage \cite{wallaceHealthInsuranceEffects2016}. 

These disparities cannot be fully explained by insurance type alone; factors such as socioeconomic status, education, and healthcare infrastructure availability play significant roles. Private health coverage often enables better access to preventive services and diagnostic care \cite{jiangImpactSupplementaryPrivate2020a}, but more research is needed to systematically explore the complex interplay between different insurance types and persistent inequities in healthcare access, quality, and long-term health outcomes.

To explore these dynamics, we focuse on three major U.S. cities: Chicago, New York City, and Washington, D.C. By analyzing neighborhood-level data, we examine multiple factors potentially associated with lead vulnerability, including water quality metrics, socioeconomic indicators, housing conditions, and health coverage types. Our analysis reveals that health coverage type demonstrates a strong correlation with lead prevalence, surpassing even well-documented factors such as housing conditions. We examine the relationship between these coverage patterns and lead prevalence, defined as the number of elevated BLL cases per 1,000 tests conducted in children who received blood lead testing. Beyond identifying correlations, we develop the Priority Score, which integrates lead prevalence, the proportion of untested children, and reliance on public health coverage to systematically rank neighborhoods for intervention. This comprehensive scoring framework enables more effective and equitable resource allocation for blood lead testing programs and public awareness campaigns by accounting for both immediate risk and systemic vulnerability. Large language models (LLMs) have shown promise across diverse domains \cite{fei2024lawbench, afane2024next,rillig2023risks, meyer2023chatgpt}, and have specifically garnered attention for their potential in supporting public health decision-making, with recent work anticipating their application to resource allocation and intervention planning \cite{qin2025opportunities}.While some studies have explored LLM deployment for public health applications \cite{jo2023understanding, thirunavukarasu2023large,haltaufderheide2024ethics,yang2023large, lupton2024generative}, empirical evaluation of their resource allocation capabilities in real-world scenarios remains limited. Building on this gap, we investigate whether current LLMs with agentic reasoning and deep research capabilities can effectively allocate limited public health resources when presented with resource allocation scenarios.
Figure~\ref{fig:v1} illustrates the alignment between model recommendations and empirical vulnerability rankings for Chicago neighborhoods, highlighting where models correctly identified high-priority areas and where they allocated resources to lower-priority neighborhoods. This visualization reveals how these advanced agentic models fall short of their advertised capabilities.

The main contributions of this study are:
\begin{itemize}
    \item Analysis of 136 neighborhoods across Chicago, New York City, and Washington, D.C., revealing that health coverage type demonstrates a strong correlation with lead prevalence among examined factors, including housing conditions, water quality, and socioeconomic factors.
    \item Development of a Priority Score that integrates lead prevalence, testing gaps, and health coverage patterns to systematically identify and prioritize neighborhoods requiring urgent intervention.
    \item Evaluation of state-of-the-art LLMs with agentic reasoning and deep research capabilities on resource allocation tasks, revealing significant limitations as models frequently overlook high-priority neighborhoods while allocating disproportionate resources to lower-priority neighborhoods, highlighting critical gaps that need to be addressed before real-world deployment.
\end{itemize}

Unlike previous research that primarily identified areas with elevated BLL, this study highlights specific neighborhoods within these high-risk areas that may be even more vulnerable due to their reliance on public health coverage, which demonstrates a strong correlation with lead vulnerability. The Priority Score provides a systematic framework for targeting interventions, while the LLM evaluation reveals critical gaps in current artificial intelligence (AI) capabilities for equitable public health resource allocation.

\section{Related Work}

Research on lead exposure has predominantly focused on understanding its health effects. Extensive literature documents the severe impacts of lead on neurodevelopment and cognitive abilities \cite{hauptmanUpdateChildhoodLead2017,canfieldIntellectualImpairmentChildren2003}. These studies have been instrumental in shaping policies around acceptable exposure thresholds and screening protocols, particularly for children who face disproportionate risks due to their developmental vulnerability.

A substantial body of work has examined the geographic and demographic patterns of lead exposure in urban environments. Prior research has primarily focused on identifying socioeconomic disparities \cite{gleasonDrinkingWaterLead2019,jacobsPrevalenceLeadbasedPaint2002}, with studies consistently demonstrating that low-income neighborhoods face elevated exposure risks. Racial inequalities have also been extensively documented \cite{moodyRacialGapChildhood2016}, revealing systematic patterns where minority communities experience higher BLLs and reduced access to testing and remediation services. Housing-related risks \cite{spanierContributionHousingRenovation2013} are major contributors to lead exposure in cities, particularly in areas with older housing stock containing lead-based paint and deteriorating infrastructure. Environmental factors, including contamination in water systems \cite{pocockEffectsTapWater1983}, have also been highlighted as significant pathways for childhood lead exposure.

Despite these well-documented contributors, limited attention has been given to how health coverage type influences lead vulnerability, mitigation strategies, and testing availability. Most existing research treats health insurance status as a binary variable (insured vs. uninsured) or as a covariate in broader socioeconomic analyses, rather than examining the distinct patterns associated with public versus private coverage. This gap is particularly significant given that health coverage type in the U.S. may serve as a more readily available and scalable indicator of community-level vulnerability, especially in areas where direct lead testing data remain incomplete. Recent advances in AI have prompted exploration of LLMs for public health applications \cite{qin2025opportunities}. While optimism surrounds their potential for supporting intervention planning and resource allocation, empirical evaluation of their capabilities remains limited. Jo et al. \cite{jo2023understanding} examined LLM deployment in public health settings, identifying both benefits and significant challenges in supporting population health needs. Among recent developments, deep research modes have emerged as a promising advancement in agentic AI, designed to autonomously gather information and synthesize comprehensive analyses \cite{xu2025comprehensive}. However, rigorous testing of their performance on real-world resource allocation tasks remains scarce. Our work extends this line of inquiry by evaluating whether LLMs can effectively perform resource allocation for real-world tasks that require balancing multiple risk factors and making equitable distribution decisions under resource constraints.

\section{Methodology}

\subsection{Data Sources and Prevalence Definition}

Throughout this work, lead prevalence refers to the number of children with elevated BLL per 1,000 tests conducted in a given neighborhood. This metric quantifies the proportion of tested children who exceed the established BLL threshold, serving as a key indicator of lead exposure risk within communities. While some jurisdictions report total tests and total cases separately, it is standard practice to express this relationship as cases per 1,000 tests, which we use consistently across all three cities for comparative analysis.

This study uses publicly available data from a range of credible sources to investigate the prevalence of elevated BLL, health insurance coverage, and their relationship to lead-related risks. For Chicago, data was obtained from the Illinois Healthy Homes and Lead Poisoning Prevention Surveillance System (HHLPSS), maintained by the Illinois Department of Public Health. This dataset provides detailed information on BLL prevalence, case counts, and community-level demographic attributes \cite{ChildhoodLeadPoisoninga}. At the time of this work, HHLPSS continued to use the Illinois threshold of 5 micrograms per deciliter (\(5 \, \mu g/dL\)) for elevated BLL. New York City data was derived from the 2024 report submitted by the New York City Department of Health and Mental Hygiene (DOHMH) to the City Council \cite{LeadPoisoningReports}. This report includes lead testing results, and testing compliance. The threshold for elevated BLL in this report remained at \(5 \, \mu g/dL\) during the study period.

For Washington, D.C., lead exposure and screening data were extracted from the "Childhood Lead Screening Report, Fiscal Year 2022," produced by the Department of Energy and Environment (DOEE) \cite{LeadDistrictDoee}. This report offers insight into screening compliance, trends in elevated BLL, and their demographic and geographic distribution. The lead prevalence in Washington, D.C., was reported using ZIP code-level data from the DOEE report. Unlike Chicago and New York, Washington, D.C., adopted the revised CDC threshold of \(3.5 \, \mu g/dL\) for elevated BLL at the time of the study.

Health insurance coverage data for all three cities was retrieved from the publicly available American Community Survey dataset provided by the U.S. Census Bureau \cite{CensusBureauData}. 

\subsection{Analysis and LLM Evaluation Framework}

We analyze neighborhood-level data to identify factors most strongly correlated with lead vulnerability, examining water quality metrics (Residual Free Chlorine and Average Turbidity), socioeconomic indicators, housing conditions, and health coverage types. Our correlation analysis reveals that health coverage patterns demonstrate stronger associations with lead prevalence than traditional risk factors, with public health coverage showing positive correlations ranging from 0.41 to 0.63 across cities. Building on these findings, we develop a Priority Score that combines lead prevalence, testing gaps, and public health coverage reliance to systematically rank neighborhoods for intervention. This framework enables public health agencies to target resources toward communities facing both documented exposure and structural vulnerability.

To evaluate whether LLMs can support resource allocation decisions in public health contexts, we assess multiple state-of-the-art models including ChatGPT, Gemini, and Claude using multiple configurations of their agentic reasoning and deep research modes. We provide each model with a structured prompt describing a realistic resource allocation scenario involving the distribution of 1,000 blood lead test kits across neighborhoods in each city. Details of the prompt design and evaluation results are provided in Section~\ref{sec:llm_evaluation}.

\section{Results}

Our correlation analysis examined multiple factors associated with lead vulnerability across 136 neighborhoods in Chicago, New York City, and Washington, D.C. Table~\ref{tab:correlation_factors} summarizes the correlation coefficients for various risk factors with lead prevalence. Water quality showed relatively weak correlations, while socioeconomic factors such as median income exhibited negative correlations. Housing conditions demonstrated particularly strong correlations among traditional risk factors, with renter-occupied homes, peeling paint, and structural problems all showing values at or above 0.59.

\begin{table}[h]
\centering
\caption{Correlation of Risk Factors with Lead Prevalence}
\label{tab:correlation_factors}
\begin{tabular}{lc}
\hline
\textbf{Risk Factor} & \textbf{Correlation} \\
\hline
\multicolumn{2}{l}{\textit{Water Quality Metrics}} \\
\quad Average Residual Free Chlorine & 0.27 \\
\quad Average Turbidity & 0.50 \\
\hline
\multicolumn{2}{l}{\textit{Socioeconomic Factors}} \\
\quad Median Income & -0.24 \\
\hline
\multicolumn{2}{l}{\textit{Housing Conditions}} \\
\quad Renter-Occupied Homes & 0.62 \\
\quad Homes with Peeling Paint & 0.59 \\
\quad Homes with Leaks & 0.57 \\
\quad Three or More Housing Problems & 0.54 \\
\hline
\end{tabular}
\end{table}

While these correlations provide valuable insights into lead vulnerability factors, the analysis of health coverage patterns reveals an even more significant relationship. Public health coverage demonstrated correlations ranging from 0.41 to 0.63 across all three cities, exhibiting both strength and consistency comparable to housing conditions. However, housing have been the subject of extensive research \cite{clarkConditionTypeHousing1985, jacobsEnvironmentalHealthDisparities2011, ScreeningHousingPrevent}, whereas health coverage patterns remain an under-explored indicator of lead vulnerability. Given this gap in the literature and the strength of the observed correlations, this work focuses on examining both the association between health coverage and existing lead prevalence, as well as the potential for predicting risk in untested areas using coverage data. 

\begin{figure}[!h]
    \centering
    \includegraphics[width=\linewidth]{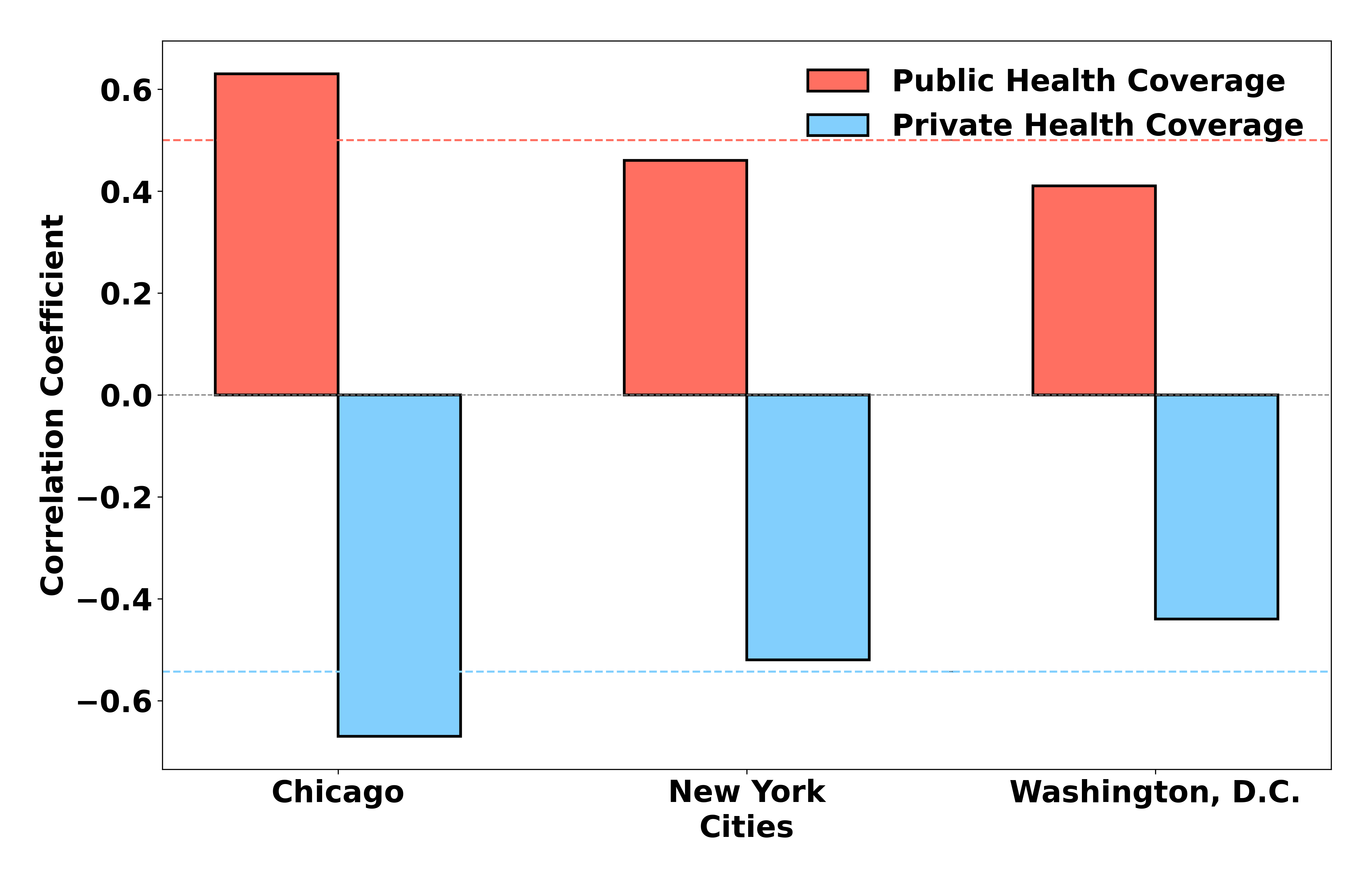}
    \caption{Correlation of Vulnerability with Health Coverage Types across Chicago, New York City, and Washington, D.C. Public health coverage is generally associated with higher vulnerability indices (average correlation: 0.50), while private health coverage shows an inverse correlation (average correlation: -0.54), as indicated by the dotted lines.}
    \label{fig:corr_health_coverage}
\end{figure}

Figure~\ref{fig:corr_health_coverage} reveals the detailed relationship between health coverage types and exposure risks across the three cities analyzed. Neighborhoods with predominantly public health coverage exhibit higher correlations with indicators of lead exposure, suggesting that these areas face significant systemic inequities in addressing environmental health risks. In Chicago, public health coverage shows a strong positive correlation of 0.63 with lead exposure metrics, while private health coverage is associated with a significant negative correlation of -0.67, reflecting a stark disparity in risk mitigation between the two coverage types. In New York City, the correlation with public health coverage is 0.46, and with private health coverage, it is -0.52, highlighting similar trends but with slightly lower magnitudes. Washington, D.C. exhibits the lowest correlations of 0.41 for public health coverage and -0.44 for private health coverage, suggesting that disparities persist but the gap is narrower compared to the other cities.

These findings highlight the structural challenges faced by neighborhoods with higher reliance on public health coverage, where limited resources and infrastructure often fail to adequately address lead exposure risks. The persistence of these disparities across distinct urban contexts further underscores how inequitable investment patterns and limited service capacity reinforce exposure vulnerability. The negative correlations observed with private health coverage across all three cities underscore its association with reduced risks, likely due to greater access to preventive healthcare, housing maintenance, environmental remediation, and educational outreach. Overall, this analysis emphasizes the urgent need for sustained, data-driven interventions in neighborhoods with high public health coverage rates to address systemic inequities and expand access to comprehensive lead exposure prevention measures.

\begin{figure*}[!t]
    \centering
    \begin{subfigure}[t]{0.32\textwidth}
        \centering
        \includegraphics[width=\textwidth]{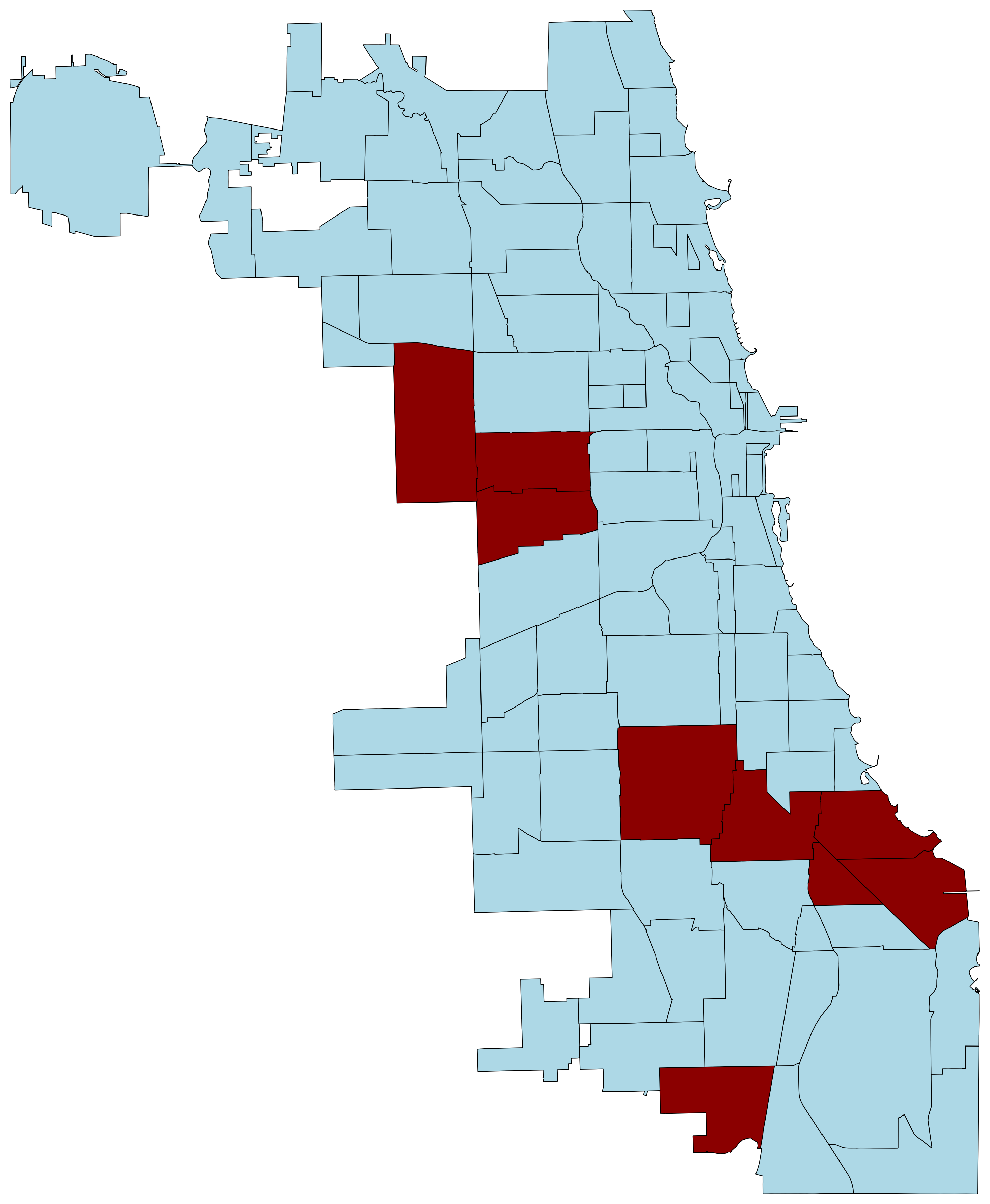}
        \caption{Chicago}
        \label{fig:chicago}
    \end{subfigure}%
    \hfill
    \begin{subfigure}[t]{0.32\textwidth}
        \centering
        \includegraphics[width=\textwidth]{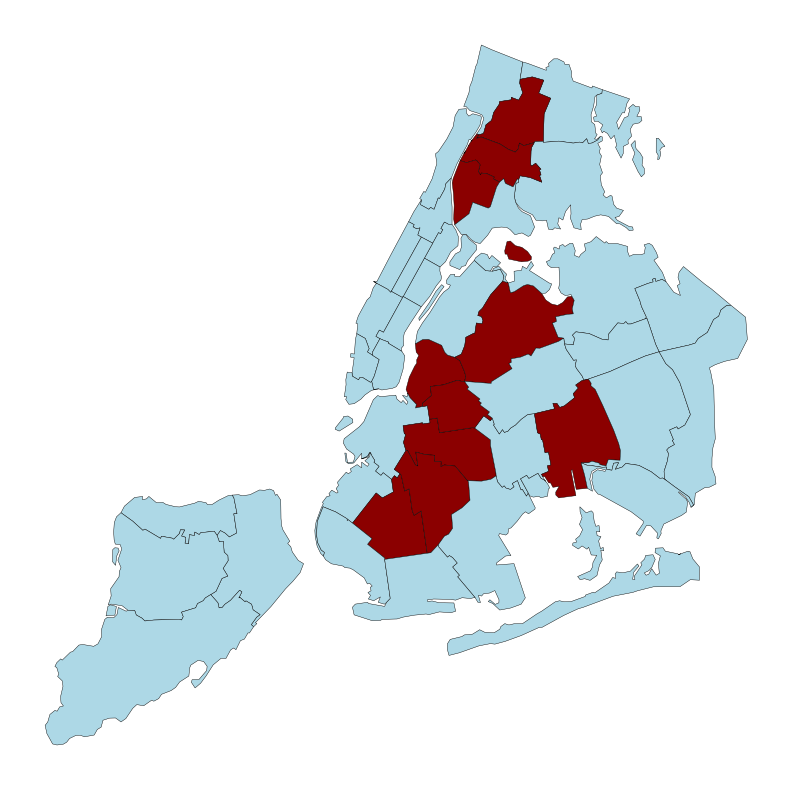}
        \caption{New York City}
        \label{fig:newyork}
    \end{subfigure}%
    \hfill
    \begin{subfigure}[t]{0.32\textwidth}
        \centering
        \includegraphics[width=\textwidth]{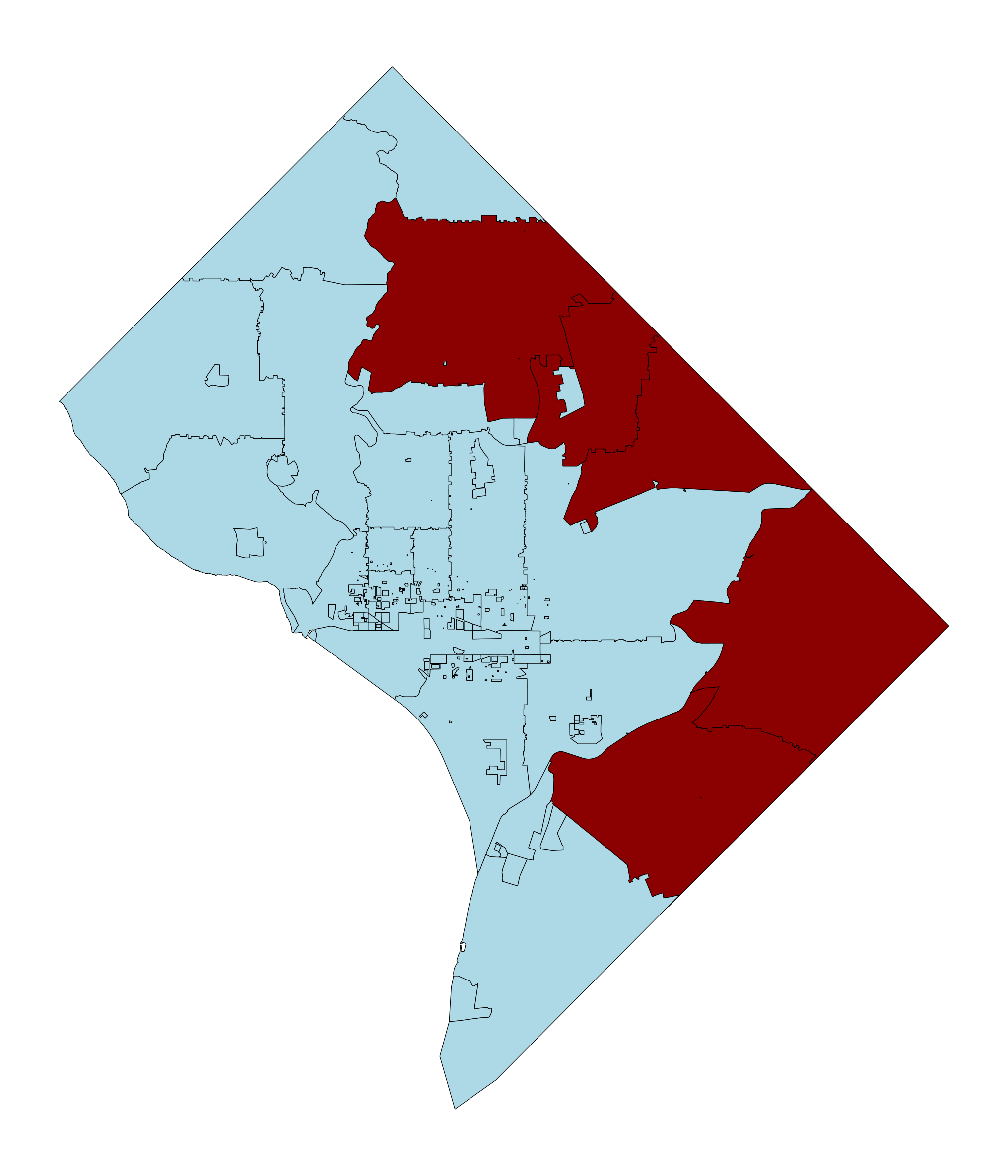}
        \caption{Washington D.C.}
        \label{fig:dc}
    \end{subfigure}

    \caption{Maps of Chicago, New York City, and Washington D.C., showing neighborhoods with the highest Priority Score values for targeted interventions. These highlighted areas represent regions with the greatest need for enhanced lead testing and remediation efforts, particularly in neighborhoods with a high concentration of public health coverage.}
    \label{fig:maps}
\end{figure*}
\subsection{Health Coverage Dynamics Across Cities}

\textbf{Chicago:} The 9 neighborhoods with a prevalence value of 4 or more, representing the proportion of children tested who were found to have BLLs exceeding the threshold of 5 \(\mu g/dL\), had an average public health coverage of 37\%, compared to the city average of 24.4\%. In contrast, neighborhoods with a prevalence value of less than 1 had an average public health coverage of 15.4\%, demonstrating a clear inverse relationship between prevalence and private insurance access. This includes 20 neighborhoods such as Lincoln Park, the Loop, and North Center. 
Two neighborhoods with the highest prevalence values, West Englewood and Englewood stand out significantly, with both having a public health coverage rate of 46\%, among the highest in the city. These neighborhoods also exhibit a sharp contrast in testing adequacy and resource availability compared to lower-prevalence areas across the city.

\textbf{New York City:} A similar pattern is observed in New York City, where neighborhoods with a prevalence value exceeding 10 average 32\% of residents with public health coverage. In contrast, neighborhoods with prevalence values below 3 have an average public health coverage of just 10\%. Two notable outliers include East Harlem, with 42\% of residents publicly covered and a prevalence rate of 3.6, and Greenpoint, which consistently has the highest prevalence rates in the city despite only 22\% of residents being publicly covered. Other neighborhoods at both ends of the prevalence spectrum appear to align with these general trends, emphasizing the connection between higher prevalence rates and increased reliance on public health coverage.

\textbf{Washington, D.C.:} In Washington, D.C., neighborhoods with a prevalence value exceeding 20 had an average public health coverage of 36\%, while those with values below 10 averaged 18.3\%. The correlation between public health coverage and prevalence appeared somewhat weaker compared to the other cities analyzed. Notably, ZIP code 20032 in the southern part of the city and ZIP code 20019 in the eastern region both have public health coverage rates exceeding 40\% despite prevalence values remaining below the city average.

This analysis highlights the distinct disparities in health coverage and lead exposure risk across cities, emphasizing the need for targeted interventions in neighborhoods with higher prevalence rates and higher reliance on public health coverage.

\subsection{Priority Score Algorithm}

While prevalence rates and gaps in testing are crucial indicators for identifying vulnerable neighborhoods, this study also incorporates public health coverage as a key factor, recognizing its strong correlation with heightened vulnerability. By integrating these metrics, the developed Priority Score provides a systematic approach to rank neighborhoods based on their need for targeted interventions.

In resource allocation contexts, assigning model weights based on domain expertise has proven effective, particularly in situations requiring nuanced decision-making \cite{rahmanAmplifyingDomainExpertise2020}. This approach ensures that the Priority Score reflects practical and context-specific considerations, enhancing its relevance for targeted resource distribution. The integrates three metrics: Prevalence of elevated BLLs (\(P\)), Percentage of untested children (\(U\)), Proportion of residents with public health insurance (\(H\)). 

The formula for calculating the Priority Score is given as:
\[
PS_n = \alpha \cdot P_n + \beta \cdot U_n + \gamma \cdot H_n,
\]

where:
\begin{itemize}
    \item \(P_n\): Prevalence of elevated BLLs in neighborhood \(n\),
    \item \(U_n\): Percentage of untested children in neighborhood \(n\),
    \item \(H_n\): Public health coverage ratio in neighborhood n,
    \item \(\alpha, \beta, \gamma\): Weights assigned to each metric.
\end{itemize}

\begin{algorithm}[h]
\caption{Priority Score Calculation for Neighborhood Ranking}
\label{alg:priority_score}
\begin{algorithmic}[1]
\Require Neighborhood data with prevalence $P$, untested-children percentage $U$, and public-coverage percentage $H$
\Ensure Ranked list of neighborhoods based on Priority Score $PS$

\State $PriorityList \gets [\,]$
\For{each neighborhood $n$}
  \State $\gamma \gets \mathrm{corr}\!\left(H,\ \text{lead prevalence}\right)$
  \State $\beta \gets (1-\alpha)(1-\gamma)$
  \State $PS_n \gets \alpha P_n + \beta U_n + \gamma H_n$
  \State append $(n, PS_n)$ to $PriorityList$
\EndFor
\State sort $PriorityList$ in descending order of $PS$
\State \Return $PriorityList$
\end{algorithmic}
\end{algorithm}

\begin{table*}[t]
\centering
\caption{Priority Neighborhoods for Increased Lead Testing Efforts Across Cities}
\label{tab:priority_neighborhoods}
\small
\begin{tabular}{l l c c p{7cm}}
\toprule
\textbf{City} & \textbf{Neighborhood/ZIP} & \textbf{PS} & \textbf{Public Health Coverage (\%)} & \textbf{Comments} \\ 
\midrule
Chicago & Englewood & 10.0 & 46 & High prevalence, large number of untested children. \\ 
Chicago & West Englewood & 9.94 & 46 & High prevalence, high public health coverage. \\ 
Chicago & Austin & 9.02 & 37 & High prevalence, large number of untested children. \\ 
Chicago & South Shore & 8.47 & 42 & Extremely large number untested children. \\ 
Chicago & Gr. Grand Crossing & 8.26 & 46 & High prevalence, high public health coverage. \\ 
\midrule
New York City & Borough Park & 10.0 & 40 & High prevalence, large number of untested children. \\ 
New York City & Bedford Stuyvesant & 9.04 & 32 & High prevalence, large number of untested children. \\ 
New York City & Greenpoint & 8.91 & 22 & High prevalence, average public health coverage. \\ 
New York City & West Queens & 8.22 & 35 & High public health coverage and untested children. \\ 
\midrule
Washington, D.C. & 20011 & 10.0 & 30 & High prevalence, large number of untested children. \\ 
Washington, D.C. & 20020 & 6.10 & 56 & High prevalence, high public health coverage. \\ 
Washington, D.C. & 20002 & 4.10 & 42 & High prevalence, large number of untested children. \\ 
\bottomrule
\end{tabular}
\end{table*}

The weight \(\alpha\) for prevalence is fixed at 0.5, reflecting its importance as a primary indicator of lead exposure risks.  The remaining weights, \(\beta\) and \(\gamma\), are dynamically determined based on their respective correlations with lead prevalence. For instance, if the correlation between public health coverage and lead prevalence is 0.6, then \(\gamma = 0.6 \times (1 - \alpha)\), ensuring that the weight for public health coverage reflects its empirical significance. Similarly, \(\beta = (1 - \alpha) \times (1 - \gamma)\), adjusting to account for the relative influence of untested children. The detailed implementation of this framework, including the calculation of the Priority Score, is outlined in Algorithm~\ref{alg:priority_score}. 
This methodology provides a structured framework for identifying neighborhoods with the highest priority for interventions, while also encouraging a more focused approach on families within these neighborhoods who rely on public health coverage. These families are often at greater risk due to systemic inequities, and addressing their specific needs can significantly improve the efficiency and effectiveness of interventions. By narrowing the scope to include the most vulnerable sub-communities, the Priority Score not only guides resource allocation but also ensures that public health actions have a more meaningful and targeted impact in mitigating lead exposure risks.

Figure~\ref{fig:maps} visually represents neighborhoods with the highest Priority Scores across Chicago, New York City, and Washington, D.C. It highlights the ten neighborhoods with the highest Priority Scores in Chicago and New York City, and the five neighborhoods with the highest Priority Scores in Washington, D.C., providing a geographic overview of areas requiring targeted interventions. This visualization allows for a broader understanding of where disparities in lead testing and health coverage are most pronounced, emphasizing the most critical areas for public health action. Notably, a clear geographical pattern emerges, with clusters of high-vulnerability neighborhoods likely driven by a combination of increased exposure risks, and reliance on public health coverage, and under-testing.

In contrast, Table~\ref{tab:priority_neighborhoods} provides a detailed description of these neighborhoods, including specific metrics and concise comments on their vulnerability. For instance, in Chicago, neighborhoods such as Gr. Grand Crossing, West Pullman, and South Lawndale are highlighted as critical areas for intervention. Similarly, in New York City, areas including East Flatbush - Flatbush and Fordham - Bronx Park stand out due to their high Priority Scores and associated disparities. In Washington, D.C., ZIP codes 20017 and 20018 are included among areas with significant vulnerabilities.

\section{LLM Resource Allocation Analysis}
\label{sec:llm_evaluation}

\begin{table*}[t]
\centering
\caption{LLM resource allocation priorities across cities (top three neighborhoods per model)}
\label{tab:llm_allocation}
\small
\begin{tabular}{l l c l c l c}
\toprule
\textbf{Model} & \multicolumn{2}{c}{\textbf{Chicago}} & \multicolumn{2}{c}{\textbf{New York City}} & \multicolumn{2}{c}{\textbf{Washington, D.C.}} \\
\cmidrule(lr){2-3} \cmidrule(lr){4-5} \cmidrule(lr){6-7}
& \textbf{Neighborhood} & \textbf{Tests} & \textbf{Neighborhood} & \textbf{Tests} & \textbf{ZIP Code} & \textbf{Tests} \\
\midrule
\multirow{3}{*}{ChatGPT 5 (Deep research)} & Austin & 150 & Borough Park & 170 & 20020 & 180 \\
& Englewood & 110 & Williamsburg – Bushwick & 135 & 20011 & 180 \\
& North Lawndale & 90 & Hunts Point – Mott Haven & 120 & 20002 & 140 \\
\cmidrule{2-7}
\multirow{3}{*}{ChatGPT 5 (Agent mode)} & South Lawndale & 120 & Greenpoint & 49 & 20020 & 110 \\
& Austin & 112 & Williamsburg - Bushwick & 31 & 20011 & 107 \\
& Englewood & 96 & East New York & 30 & 20019 & 85\\
\cmidrule{2-7}
\multirow{3}{*}{CLaude 4.5 (Deep research)} & Austin & 100 & Greenpoint & 120 & 20032 & 150 \\
& West Garfield Park  & 75 & Borough Park & 100 & 20019 & 150 \\
& Humboldt Park & 70 & Hunts Point-Mott Haven & 90 & 20010 & 100 \\
\cmidrule{2-7}
\multirow{3}{*}{CLaude 4.5 (Extended thinking)} & West Garfield Park & 100 & Bedford-Stuyvesant & 100 & 20011 & 120 \\
& Austin & 100 & Mott Haven & 80 & 20020 & 100 \\
& Englewood  & 100 & East Harlem  & 80 & 20032 & 100  \\
\cmidrule{2-7}

\multirow{3}{*}{Gemini 2.5 Flash (Deep research)} & North Lawndale & 112 & Hunts Point - Mott Haven & 95 & 20032 & 255 \\
& Englewood & 112 & High Bridge - Morrisania & 88 & 20019 & 250 \\
& West Garfield Park & 107 & Crotona - Tremont & 79 & 20002 & 150 \\
\bottomrule
\end{tabular}
\end{table*}

We evaluate the resource allocation capabilities of advanced LLMs that operate in their research-augmented and agentic modes, assessing how effectively they distribute limited public health resources under multi-factor constraints. Modern LLMs such as ChatGPT, Gemini and Claude now include expanded reasoning modes that simulate complex policy analysis workflows. These modes differ substantially in architecture and execution: \textit{Deep Research} and \textit{Extended Thinking} allow models to autonomously search and synthesize information from dozens of external sources, often running for extended periods of time (up to 30 minutes) to build detailed, evidence-based responses. By contrast, \textit{Agent Mode} executes reasoning chains interactively, invoking multiple subroutines to simulate stepwise problem solving. These capabilities are designed to emulate systematic research behavior and, in theory, enable comprehensive assessments of multi-dimensional risk factors that would otherwise require expert domain knowledge.

To test whether these capabilities translate into meaningful decision-making accuracy, we evaluated four configurations across three major urban contexts:

\begin{enumerate}[label=(\arabic*)]
    \item ChatGPT 5 (Deep Research)
    \item ChatGPT 5 (Agent Mode)
    \item Claude 4.5 (Deep Research)
    \item Claude 4.5 (Extended Thinking)
    \item Gemini 2.5 Flash (Deep Research)
\end{enumerate}

Each model was provided with a standardized resource allocation task designed to measure its ability to prioritize high-risk neighborhoods for blood lead testing based on epidemiological and socioeconomic indicators with a focus on its top 3 recommendations. The prompt below was used uniformly across all systems to ensure methodological consistency.

\begin{tcolorbox}[colback=gray!8, colframe=gray!50, boxrule=0.5pt, arc=2mm, left=5mm, right=5mm, top=3mm, bottom=3mm]
\textit{Prompt:} We have 1,000 blood lead test kits for children to distribute across [City] neighborhoods. Conduct deep research to identify and rank neighborhoods by priority based on: (1) number of untested children, (2) historical lead exposure prevalence, (3) typical elevated blood lead level rates per 1,000 tests conducted, and (4) percentage of residents with public health coverage, a strong indicator of community vulnerability and limited access to preventive healthcare. Provide a priority-ordered list with recommended resource allocation per neighborhood.
\end{tcolorbox}

Table~\ref{tab:llm_allocation} summarizes the top three neighborhoods identified by each model for blood lead test kit distribution, based on a total of 1,000 kits per city.

\subsection{Evaluation Across Cities}

\paragraph{Chicago.} 
The highest-priority neighborhoods in Chicago, based on both our Priority Score calculations and historical lead data, are \textbf{Englewood}, \textbf{West Englewood}, and \textbf{Austin}. These areas consistently exhibit the highest prevalence of elevated BLLs and the largest populations of untested children. Despite this clear empirical evidence, the LLMs did not prioritize West Englewood in their top three allocations. Englewood itself appeared in a few model outputs but generally received fewer resources than warranted by its vulnerability, typically ranked around fourth to sixth. Instead, models like ChatGPT 5 (Agent mode), Gemini, and Claude 4.5 (Extended thinking) allocated significant resources to areas such as South Lawndale and West Garfield Park, both of which exhibit comparatively lower prevalence. This highlights the LLMs’ difficulty integrating multiple vulnerability indicators and weighting lead prevalence appropriately under resource constraints.

\paragraph{New York City.} 
According to prior public health reports and our Priority Score framework, \textbf{Borough Park}, \textbf{Greenpoint}, and \textbf{Bedford-Stuyvesant} consistently rank as the city's highest-priority neighborhoods for lead testing. While these areas appeared in several model recommendations, they frequently received lower allocations than warranted by their risk levels. ChatGPT 5 (Deep research), Gemini, and Claude 4.5 (Deep research) allocated considerable resources to Hunts Point - Mott Haven, placing it third despite historic data indicating very low lead prevalence compared to top-priority areas. Among the evaluated models, Claude 4.5 (Deep research) most closely aligned with empirical vulnerability rankings, correctly prioritizing Greenpoint and Borough Park in the first two spots. Despite the abundance of publicly available data models still made notable prioritization errors, suggesting that data availability alone does not guarantee accurate allocation decisions.

\paragraph{Washington, D.C.} 
For Washington, D.C., ZIP codes \textbf{20011}, \textbf{20020}, and \textbf{20002} have the highest documented lead prevalence (even without factoring in the health coverage data introduced in this paper). ZIP code 20011 has almost twice the lead prevalence of the next highest area, making it the most critical testing priority. Despite this, only Claude 4.5 (Extended thinking) assigned 20011 the highest allocation, while Claude 4.5 and Gemini (Deep research) omitted it from the top three altogether. ChatGPT 5 (Deep research) and ChatGPT 5 (Agent mode) both placed 20011 second, demonstrating limited consistency in prioritization. Several models overemphasized ZIP codes with moderate prevalence but higher population density, reflecting a pattern of heuristic bias that favors visible demographic factors over measured exposure risk. 

\paragraph{Summary.}
Model outputs across all three cities reveal recurring misalignments between empirical vulnerability indicators and recommended allocations. To quantify this, we compute an \textit{accuracy} metric measuring whether a model's top three prioritized neighborhoods include the empirically highest-risk areas for each city, regardless of internal ranking. The target high-risk neighborhoods are Chicago: Englewood, West Englewood, Austin; New York City: Greenpoint, Borough Park, Bedford-Stuyvesant; and Washington, D.C.: 20011, 20020, 20002. This metric captures whether models correctly identify key areas of concern. We also report average run time in minutes for each reasoning mode.

\begin{table}[t]
\centering
\caption{Average model run time and accuracy}
\label{tab:llm_accuracy}
\small
\begin{tabular}{l c c}
\toprule
\textbf{Model} & \textbf{Mean time (min)} & \textbf{Accuracy} \\
\midrule
ChatGPT 5 (Deep research) & 19.60 & \textbf{0.66} \\
ChatGPT 5 (Agent mode) & 23.00 & 0.55 \\
Claude 4.5 (Deep research) & 11.83 & 0.33 \\
Claude 4.5 (Extended thinking) & 2.56 & 0.55 \\
Gemini 2.5 (Deep research) & 3.12 & 0.22 \\
\bottomrule
\end{tabular}
\end{table}
Table~\ref{tab:llm_accuracy} presents the comparative evaluation results, summarizing average processing time and accuracy across all LLM configurations. The results indicate that longer autonomous reasoning workflows do not necessarily yield better prioritization accuracy. ChatGPT 5 Deep Research achieved the highest overall accuracy at 0.66, while Claude 4.5 Extended Thinking matched ChatGPT 5 Agent Mode's accuracy despite operating approximately nine times faster. These findings suggest that extended processing time alone does not translate into improved decision quality and that model reasoning depth may depend more on how contextual cues are integrated than on total execution duration.

Beyond accuracy metrics, model outputs reveal systematic failures in prioritizing well-documented high-risk areas. West Englewood in Chicago was excluded from all top-three allocations despite being among the city's most vulnerable neighborhoods. In Washington, D.C., ZIP code 20011, which records nearly double the lead prevalence of any other area, was omitted entirely by Gemini and Claude 4.5 Deep Research while receiving reduced allocations from both ChatGPT 5 variants. These omissions underscore critical limitations in current retrieval-augmented reasoning, where models fail to weight empirical vulnerability data appropriately when making allocation decisions.

\section{Discussion and Recommendations}

The findings of this study reveal systematic patterns in how health coverage type correlates with lead vulnerability and demonstrate the value of integrating socioeconomic indicators into resource allocation frameworks. Rather than representing a direct cause of lead vulnerability, health coverage type functions as an intermediate indicator that signals broader socioeconomic and infrastructural inequities constraining access to safe housing, preventive care, and remediation resources. The proposed Priority Score offers a practical framework for identifying communities requiring urgent intervention by combining health coverage data with empirical evidence of testing gaps and lead prevalence, while also providing a testbed for evaluating LLM-based resource allocation capabilities. Assessing whether advanced AI systems can translate this structured vulnerability framework into appropriate distribution decisions reveals critical limitations in current agentic reasoning modes. Key observations for LLMs applications and public health policy emerge from this analysis:

\textbf{Inaccurate Data Retrieval and Outdated Information:} 

Despite the technical promise of Deep Research and Agentic modes, which are designed to autonomously gather information from numerous sources over extended periods, models frequently cited outdated information despite the availability of more current data within the same sources. This pattern reveals fundamental limitations in how these systems prioritize and synthesize information, often failing to identify the most recent and relevant public health data even when conducting supposedly comprehensive research.
    
\textbf{Susceptibility to Non-Empirical Sources:} Models demonstrated concerning reliance on articles discussing neighborhood vulnerability without empirical support, allowing non-data-driven narratives to influence prioritization decisions. The gap between surface-level textual coherence and underlying analytical validity underscores that current retrieval-augmente reasoning capabilities cannot reliably distinguish between evidence-based analysis and speculative commentary, leading to inaccurate resource allocation recommendations.

\textbf{Health Coverage as a Vulnerability Indicator:} Health departments should systematically track health coverage status for all children with elevated BLL to enable more precise assessment of how insurance type influences exposure risk, testing access, and remediation outcomes. Given the strong correlations observed between public health coverage and lead prevalence (ranging from 0.41 to 0.63 across cities), this data represents a readily available and scalable indicator that can identify vulnerable communities, particularly in areas where comprehensive lead testing data are unavailable or limited. Interventions in high-risk neighborhoods identified through Priority Score analysis should employ targeted approaches that recognize intra-neighborhood variation in vulnerability, such as households with young children living in pre-1978 housing or families relying exclusively on public health coverage.

\section{Limitations and Future Work}

Our analysis of three major U.S. cities (Chicago, New York City, and Washington, D.C.) provides detailed insights into the relationship between health coverage patterns and lead vulnerability in urban environments. Extending this framework to additional cities with documented lead crises, such as Flint, Detroit, and Boston, would help determine whether the same structural patterns persist in regions facing more severe contamination. Examining smaller urban centers and communities with varying racial and economic compositions could also clarify how local housing, infrastructure, and demographic factors influence vulnerability. Beyond the United States, future work could explore whether similar disparities between socioeconomic indicators and environmental health outcomes emerge in international contexts, where differences in health systems and housing regulations may shape the dynamics of lead exposure risk. Further research should also examine the performance of other language models and emerging agentic reasoning modes to better understand where these systems fail and why. Evaluating a broader range of models would allow for comparative insights into the underlying reasoning strategies that shape allocation decisions and identify recurring blind spots in how LLMs interpret empirical data. It is also important to note that these models evolve continuously, and future versions may exhibit improved reasoning, and contextual alignment. Periodic reassessment of new model releases could therefore help track progress and reveal whether advances in reasoning depth and retrieval integration reduce the limitations observed in this study.

\section{Conclusion}

This study demonstrates that health coverage patterns serve as powerful and previously underutilized indicators of neighborhood-level lead vulnerability. Through analysis of 136 neighborhoods across Chicago, New York City, and Washington, D.C., we found that public health coverage exhibits positive correlations with lead prevalence ranging from 0.41 to 0.63, while private coverage shows negative correlations between -0.44 and -0.67. These consistent patterns across cities reveal systematic disparities in environmental health risks and healthcare access that disproportionately affect communities facing limited resources, aging infrastructure, and long-standing disinvestment. Rather than representing a direct cause of vulnerability, health coverage type functions as an intermediate indicator signaling deeper socioeconomic and infrastructural inequities that constrain access to safe housing, preventive care, and remediation resources. The Priority Score framework developed here integrates lead prevalence, testing gaps, and health coverage reliance to provide public health agencies with a structured approach for identifying neighborhoods requiring urgent intervention, moving beyond simple prevalence metrics to capture both documented exposure and structural vulnerability.

Our evaluation of large language models reveals critical limitations in their current capacity to support complex, data-driven resource allocation decisions. While ChatGPT 5 Deep Research achieved the highest accuracy of 0.66, model outputs across all configurations frequently failed to recognize the most vulnerable neighborhoods, systematically overlooking well-documented high-risk areas such as West Englewood in Chicago and ZIP code 20011 in Washington, D.C. Despite operating in Deep Research or Agentic modes designed to simulate advanced reasoning, the models demonstrated incomplete ability to connect quantitative vulnerability data with appropriate allocation decisions. In multiple cases, resources were directed toward areas with relatively lower measured risk, exposing a persistent gap between textual reasoning coherence and data-grounded prioritization. Notably, the reports generated by these LLMs often appeared detailed and well-supported, presenting seemingly accurate narratives that diverged substantially from empirical evidence when evaluated against verified vulnerability data. These findings underscore that while LLMs show potential for supporting public health resource allocation, their recommendations require careful validation and human oversight. The results emphasize the continued importance of expert judgment in guiding automated decision systems and highlight the need for substantial refinement before such tools can reliably interpret data, identify structural vulnerability, and allocate resources effectively in real-world public health contexts.

\bibliographystyle{IEEEtran}
\bibliography{references}
\end{document}